\documentstyle[proceedings,numreferences,epsfig]{crckapb}

\newcommand{\ba}{\begin{eqnarray}}
\newcommand{\ea}{\end{eqnarray}}
\newcommand{\baa}{\begin{eqnarray*}}
\newcommand{\eaa}{\end{eqnarray*}}
\newcommand{\bb}{\begin{thebibliography}}
\newcommand{\eb}{\end{thebibliography}}

\newcommand{\lab}[1]{\label{#1}}

\begin{opening}
\title{Dispersion Relations  \protect\\
        and inconsistency of $\rho$ data }
\subtitle{}

\author{O.V. SELYUGIN}
\institute{Institut de Physique, Universit$\acute{e}$ de Li$\grave{e}$ge, Belgium \\
         and  BLTPh, JINR, Dubna, Russia}

\end{opening}

\runningtitle{Dispersion Relation...        }

\begin{document}


%

\begin{abstract}
The high energy elastic nucleon cross section is treated from the viewpoint
of the basic principles of  local field theory.
 The connection between the energy
dependence of $\sigma_{tot}$ and the $\rho$ - ratio of
 the real to imaginary parts of
the forward elastic scattering amplitude is examined in the framework of 
dispersion relations,  derivative dispersion relations and  crossing symmetry.
\end{abstract}

\section{ Introduction}


  Interest in diffractive processes is now revived due, in most part,
  to the discovery of rapidity gaps at HERA, i.e. of  processes
 where diffractive interactions contribute a siseable part 
  of the whole amplitude.
The diffraction interaction  is defined by the multigluon colorless exchange
 named  pomeron \cite{landsh}. In the perturbative regime of QCD, it can be
considered as a compound system of  the two Reggeized gluons \cite{lipat}
 in the approximation where one sums the leading  $ln$'s in energy, though its
nonperturbative structure is basically unknown.
Research on the nature of pomeron requires the knowledge of the
parameters of   purely diffractive processes,  such as the total cross
section and the phase of the elastic scattering amplitude.
 These quantities are closely
related with the first principles of  field theory such as causality,
polynomial boundedness, crossing symmetry etc.
They are also important for  modern nuclear physics at high energies,
 as these quantities
 underlie  many modern descriptions of nuclear interactions
when we study such effects as nuclear shadowing, transparency or broken
parity \cite{nucl1}.

The energy dependence of the cross section and of the parameter $\rho(s)$ - the
ratio of the real to imaginary part of the scattering amplitude
 in the high energy
domain - is a much discussed question which still remains without a definite
answer.
  Quite many efforts were spent to understand the high-energy hadron scattering
from such  general principles of relativistic quantum field theory as Lorenz
invariance, analyticity, unitarity and crossing symmetry.
Analyticity, which arises from the principle of causality,
  leads to dispersion
relations which give the real part of the scattering amplitude at $t \ = \ 0$
as an energy integral involving $  \sigma_{tot}$ .

The forward dispersion relations for nucleon-nucleon scattering have not
been proven, although they were written down a long time ago \cite{gold}.
Numerous papers have been devoted to calculations of the real parts of the $pp$
and $\bar{p}p$ forward scattering amplitude, using  variety of
dispersion relations
and  different representation for the energy dependence
of the imaginary part of the scattering amplitude
 at $t=0$.

The dispersion relations for scattering amplitudes depend on three properties,
analyticity in the energy variable, the optical theorem, and polynomial
boundedness.
The prediction of the dispersion relation is very important for the discovery
of new physical phenomena at superhigh energies. For example, in the case of
potential scattering by nonlocal potentials the polynomial boundedness can be
broken  \cite{kyuri1}.
  With some additional assumptions connected with  string models
  \cite{string},
 the effect of non-local behavior will  decrease the phase, and hence
increase $\rho = ReF/ImF$.
A different behavior  can be obtained in the case of extra internal dimension
  \cite {kyuri2}.
So, such a remarkable effect ( the deviation of $\rho(s)$ from
 the prediction of
the dispersion relation) could be discovered  already at the LHC 
 in the TOTEM experiment.

\section{ Different  connections between the real and the imaginary part
             of the scattering amplitude}

The optical theorem connects the imaginary part of the
 forward elastic scattering
amplitude with the total cross section. 
$\sigma_{tot}(s)\ = \ 4 \ \pi Im F(s, t=0)$.
The crossing property of the scattering amplitude is used very often 
 to relate  the imaginary part to the real part of
 the scattering amplitude with the substitution
$  S \rightarrow \tilde{S} = S exp(-i\pi /2)$.
In the case of the maximal behavior,
the crossing-even and crossing-odd amplitudes at $t=0$ are \cite{nicled}
\ba
F_{+}(s,t=0)/is =  F_{+}(0) \ [ln(s \ e^{-i \pi /2}]^2  \nonumber \\
F_{-}(s,t=0)/is = i F_{-}(0) \ [ln(s \ e^{-i \pi /2}]^2
\ea

 On the other hand the integral dispersion relations 
give us the most powerful connections between $\sigma_{tot}$ and $\rho$.
Their form depends on the behavior of the crossing-even 
 and crossing-odd parts of the
scattering amplitude. In the case where the crossing-even part 
 of the scattering
amplitude  saturates the Froissart bound, so that the total cross section
 does not grow
 faster than a power of $\ln(s)$, the dispersion relation requires a single
subtraction. If the crossing-odd part of the scattering amplitude does not grow
with energy, so that the difference of the total cross section
between hadron-hadron
and hadron-antihadron scattering falls with energy,
the odd dispersion relation needs no  subtraction
\ba
Re f^{even}(E)  \ = \ Re F^{even}(0) + P \frac{1}{\pi} \int_{m}^{\infty} dE'
\frac{2\ E^2}{E'(E'^2-E^2)} Im f^{even}(E'); \ea
\ba
Re f^{odd}(E)  \ = \  P \frac{1}{\pi} \int_{m}^{\infty} dE' \frac{2\
E}{(E'^2-E^2)} Im f^{odd}(E'), \ea
where $P$ denotes the principal part of the integrals.

On the basis of the integral dispersion relation, one obtained
derivative dispersion relations (first used in the context of Regge theory
\cite{grib} and developed in  \cite{bronz}) which are more suitable for
calculation but, of course, have a more narrow region of validity 
of scattering amplitudes \cite{fischer1,fischer2}.
It is to be noted that in this case we lose the subtraction constant.
It means that the derivative dispersion relation
can be used     only at high energies where this constant is not perceived.
The derivative dispersion relations can be written
\ba
\frac{R(s)}{s^{\alpha}} = tan [ \frac{\pi}{2} ( \alpha -1 + \frac{d}{dln \ s})]
 \frac{I(s)}{s^{\alpha}}, \ea
where the parameter $\alpha$ is some arbitrary constant modifying the integral.
 Usually  we choose $\alpha =1$
from  the sake of simplicity \cite{fischer2}.
In \cite{menon} it was shown that if we fit the experimental
data, the best value will is  $\alpha = 1.25$. But in this case, the parameter
$\alpha \not= 1 $ will change the relation between the imaginary and
the real parts
of the scattering amplitude in the whole energy region.
So, the predictions of the integral dispersion relations
 and of the derivative dispersion relation will  not
coincide. That means that one or the other is false.
From the  viewpoint of  basic
principles, which lead to the integral dispersion relation,
 we have to give  prefer once to them. Hence, the equality of both
 representations at high energies
requires that $\alpha$  be equal to $\ 1$.

\section{ The experimental data and the check of the integral
          dispersion relations}

Now it is a common belief that the experimental data confirm the validity
  of the dispersion relations. Of course, there  is some ambiguity
  concerning the energy dependence of the total cross sections.
 Already the fit by Amaldi et al.
\cite{amaldi}  and by Amos et al. \cite{amos} has given the tendency for
$\sigma_{tot}(s)$  and $\rho(s)$ energy to increase.
Khuri and Kinoshita \cite{kyuri1} have predicted that $\rho(s)_{pp}$ should
approach zero from positive values. At the time of this prediction,
$\rho(s)_{pp}$ was known to be a small negative quantity implying, thus, that
$\rho(s)_{pp}$ should be zero at some energy, reach a maximum, and then tend to
zero at still higher energies.

The comparison of the predicted values of the real part of the $pp$ scattering
amplitude with the experimental ones allows one  to conclude that,
 on the whole,
the experimental data agree well with the dispersion relation predictions.
However if we take into account the whole set of experimental data, 
without any exception, we find  that the total $\chi^2$ is  large
and that some experimental data contradict others.
Some points of one  experiment are
situated above the theoretical line and some points of other experiment are
 below  the theoretical curve by more than three statistical
errors.
   Moreover, in \cite{e760}, it  was noted that   experimental data on
  $\bar{p}p$-scattering around $p_{L}=5 $ {\rm GeV/c} disagreed with the
  analysis  based on the dispersion relations 
  \cite{krol}.

At low energy, the definition of the subtraction constant is very important.
The spin-independent amplitude could be continued into the low-energy region
 of  $\bar{p}p$ scattering and into unphysical region.
However, because of the lack of the data for the low energy $\bar{p}p$
scattering, such a continuation (e.g.. by means of the effective range
approximation) cannot be carried out.
This is why in the dispersion relation calculations the imaginary part of the
amplitude in the unphysical and low-energy region is sometimes decomposed into
a series in which the coefficients  are determined by comparison of the
calculated real parts with the experimentally measured ones.

An alternative way of handling these regions is to replace the continuum of
states with a set of bound states at fixed energies.
Mathematically, this is equivalent to replacing  a cut of an analytic function
by a sum of poles (resonances). The values of their residues (coupling
constant) are  either a priori fixed, or are found by
comparing the calculated values of
the real part with the experimental data.
One should note that the replacement of the cut with the poles is, in itself,
an arbitrary procedure. Besides, both the exact number of poles-resonances and
the values of their coupling constants are unknown.

Therefore, it seems to us that the apparant agreement of theoretical (i.e. by
the dispersion relation) calculations with experimental data on the real parts
of the $pp$ forward scattering amplitude means only that the parameters in
various approximations of unphysical and low-energy region of $\bar{p}p$
scattering can be chosen so as to obtain consistency between the theoretical
calculations and the experimental data.
Really, the dispersion relations for $pp$ , $\bar{p}p$ forward scattering
 have been tested at low energy, at best, only quantatively.

\section{ The calculation of $\rho$ through different methods}

            As noted above, we can obtain the value of $\rho$
by  different methods.
  The simplest one is to use the crossing symmetry properties
 of the scattering amplitude and to obtain the $Re F(s,t)$
  by  changing  $ \bf S \rightarrow S exp(- i \pi /2)$
  in all energy dependent parameters of a  model, see for example \cite{nicled}.
In that case, the real
  part of the scattering amplitude  is obtained straight a way.

   Another way to obtain the $Re F(s,t)$ is to use
 the derivative  dispersion relations.
    The total cross section can be taken in form
\ba
 \sigma_{tot} = Z_{pp} +  A \ Ln(s/s_0)
          + Y_1 s^{-\eta_1}
          - Y_2 s^{-\eta_2}      \lab{stot}.
\ea
  Hereafter we suppose, beside the special separate cases, that $s$ has the coefficient
  $s_{1}^{-1} = 1 \ GeV$.
  The derivative dispersion relation gives us  \cite{compete}
\ba
 && \rho(s,t) \ \sigma_{tot}= \nonumber \\
&& \frac{\pi}{2} \ A \  ln(\frac{s}{s_0})
          - Y_1 \  s^{-\eta_1}   [\tan(\frac{1-\eta_1}{2}\pi]^{-1}
          - Y_2 s^{-\eta_2} [\cot(\frac{1-\eta_2}{2}\pi ]^{-1}. \label{ddr}
\ea
  To accurately compare the results of the different calculations,
  let us
  examine  carefully the calculations with the integral dispersion
relation.
   There were many works on the methods of  calculating
 the dispersion integrals \cite{hend1}.
      From recent  works, one should  note the paper \cite{jenkdr}
 where it was shown that
  for the most common parametrizations of the total cross section,
 the principal value integrals can be calculated in terms of rapidly convergent series. The advantages are
 not restricted to a reduction of the computer time.

  Let us consider, as in \cite{soding}, the standard dispersion relations
for $pp \rightarrow pp \ (+)$  and
 $\bar{p}p \rightarrow \bar{p}p \ (-)$
\ba
\rho(E)_{\pm} \sigma_{\pm}(E)  = \frac{B}{p}
                               + \frac{E}{\pi p} \
   P \int^{\infty}_{m} d E^{\prime} p^{\prime}
          [\frac{\sigma_{\pm}(E^{\prime})}
 {E^{\prime}(E^{\prime} -E)} - \frac{\sigma_{\mp}(E^{\prime})}
     {E^{\prime}(E^{\prime} +E)}],
\ea
 where $E$ and $p$ are the energy and the momentum in the laboratory frame, $m$ is the proton mass
 and $B$ is a subtraction constant.
 In terms of the variable $s= 2m (E+m)$,
 a parametrization of the cross section
 can be written in the form (\ref{stot}).

  In calculating the principal value of the dispersion integral,
 we face  two difficulties.
 One is the singularity of  the integral at the point $E'  =  E$
 and the other is the infinity of the upper bound
 of the integral which is especially important in the case
of a  total cross section growing  with energy.
 It is possible to solve both  problems if the integral is divided
 in three parts.
  Let us cut the integral at the singularity point and select
 the last part which contains   the tail of the integral
  \ba
  \int^{\infty}_{m} d E^{\prime} p^{\prime} [\frac{\sigma_{\pm}(E^{\prime})}
 {E^{\prime}(E^{\prime} -E)} - \frac{\sigma_{\mp}(E^{\prime})}
     {E^{\prime}(E^{\prime} +E)}] = I_1 \ + I_2 + I_3,
\ea
where
\ba
I_1= \int^{E-\epsilon}_{m} d E^{\prime} p^{\prime} [\frac{\sigma_{\pm}(E^{\prime})}
 {E^{\prime}(E^{\prime} -E)} - \frac{\sigma_{\mp}(E^{\prime})}
     {E^{\prime}(E^{\prime} +E)}],
 \ea
\ba
I_2 = \int_{E+\epsilon}^{k E} d E^{\prime} p^{\prime} [\frac{\sigma_{\pm}(E^{\prime})}
 {E^{\prime}(E^{\prime} -E)} - \frac{\sigma_{\mp}(E^{\prime})}
     {E^{\prime}(E^{\prime} +E)}],
 \ea
\ba
I_3 = \int^{\infty}_{k E} d E^{\prime} p^{\prime} [\frac{\sigma_{\pm}(E^{\prime})}
 {E^{\prime}(E^{\prime} -E)} - \frac{\sigma_{\mp}(E^{\prime})}
     {E^{\prime}(E^{\prime} +E)}].
 \ea
 Here $k$ is  a number, for example, $k=10$.
  The third integral can easily  be calculated analytically
 in some of the most important cases,
 for example, if we take the nonfalling part of the total cross section
 in the form,
\ba
 \sigma_{tot}(s)^{as} \ = \ Z \ + \  A \  ln^n (s/s_0 ).
\ea
 Here we take for simplicity $s_0=1$ and $n=1$ or $= 2$.
 The other cases are  slightly more complicated,
  but do not lead to any principal difficulty.

 Let us take $k=10$.
 For the constant term the integral is
\ba
\int_{k E }^{\infty}&& [\frac{2Z \ E}{E'(E'^{2}-E^{2})}  ] dE'  =
 2 Z \  E \ [ \frac{1}{2 k E} (ln(1+a)-ln(1-a)  \nonumber \\
 && +ln(-\frac{1}{E})
  -ln(\frac{1}{E}))]  \ = \ 2.00671 \ Z /k \  = \ 0.200671 \ Z.
\ea

 For the $n=1$ the growing term is
\ba
\int_{k E}^{\infty}[\frac{2 E \ A \ ln{ E'}}{ E'(E'^2-E^2)}  ]  dE' \  =
\   A \ [0.662285 +0.20067 \ ln E]. \lab{tail1}
\ea

 In the case of maximal, allowed by analyticity,
 growth of the total cross section, $I_3$ is given by
\ba
\int_{k E}^{\infty} [\frac{2 E A \ ln^2{ E'}}{ E'(E^2-E^2)}  ]  dE' \ =
\frac{A}{12}
 [28.6337 +  ln E \ (15.8948 + 2.40805 \ ln E)] . \lab{tail2}
\ea

  As a result, we can compare our computer calculations
 with our analytical
 calculation to find a suitable upper bound of the dispersion integral
 for computer calculation.
  Such a comparison is shown in Figs. 1 and 2
  where the line is the analytical
 calculations,  and the box and triangles represent computer calculations.
 It is clear that both  calculations coincide with high accuracy 
 in all energy region.

   Now return to the first two integrals. The numerical calculations show
  that,  after removing the tail of the whole integral,
   the sum of the  integrals $I_1$ and $I_2$ very fastly tends
  to its limit and  cancels out the divergent parts.
   Usually, a precision of 1 \% is  sufficient. 
   As a result, we can easily control the accuracy of our calculations.
The calculations of  $\rho(s)$
   by the dispersion relation are shown in Fig.1 for the case
$\sigma_{tot}(s) \approx ln (s)$
    and  in Fig.2 for the case $\sigma_{tot}(s) \approx ln^2 (s)$.

\begin{figure}[t!]
 \epsfysize=6.cm
\epsfxsize=12.cm
\centerline{\epsfbox{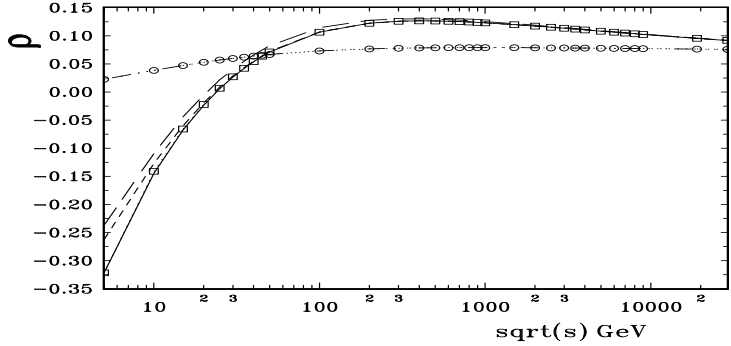}}
\caption{ The calculations of $\rho(s)$ for the energy dependence
  of $\sigma_{tot}$ given by (\ref{stot})
 with $n=1$ (the solid line - by the derivative dispersion relations (DDR) (\ref{ddr}),
  the short dashed line - by the integral dispersion relations (IDR)
  with the subtracted constant from work by Amos,
 the long dashed line - by  using the crossing symmetry,
   the squares - by the integral dispersion relations with fitting subtracting
 constant).
  In the upper part of the figure the calculations
 for the tail part of dispersion integral are presented
 ( the  line - by the analytical (\ref{tail1}) and the circle - by numerical).
 }
\end{figure}

\begin{figure}
 \epsfysize=6.cm
\epsfxsize=12.cm
\centerline{\epsfbox{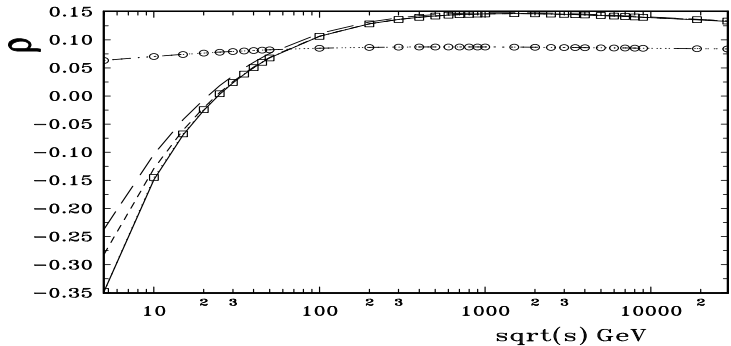}}
\caption{The same as  Fig.1 with $n=2$}
\end{figure}

       The calculations  by the derivative dispersion relation are shown
 in these figures also.
   At high energies, both calculations coincide and give the same predictions.
   If we choose the corresponding value of the  subtracting constant $B$,
 both  calculations   coincide with high accuracy
   at low energy too.
    The calculation  using  crossing-symmetry also coincides
 with  both  calculations in the
   high energy region and can coincide with the integral dispersion relations
   if we choose the corresponding
   size of the subtraction constant. But it will differ from
   the subtraction constant required for the derivative
   dispersion relations.

So, we can conclude
  that the descriptions at low and high energy of $\rho(s)$
 are practically independent of each other in the case
  of the integral dispersion relation. The subtracting constant practically
  unties these two regions.
   So, the good description of the data of $\rho$
 cannot give  valuable information about the validity of
    our basic principle and weakly impacts on our prediction
 about the behavior of the total cross section and the parameter
   $\rho(s)$ at high energies.

\section{ The experimental data on $\rho$}

    Let us now consider very shortly the procedure of extracting parameters
  of the hadronic
    scattering amplitude from the experimental data $dN/dt$.
    In fact, in experiment we measure $dN/dt$, as a result of which
      ``experimental" data such as $\sigma_{tot}$, $B$-slope and $\rho$
  are extracted  from  $dN/dt$ with some model assumptions.
  Some assumptions are also needed to extrapolate  the measured quantities 
  to $t = 0$.
 Most important in this case is the determination of the normalization
 of the experimental data.
 Contributions from the electromagnetic and hadronic interactions
 exist in the hadron interaction at small angles.
 We can calculate the Coulomb amplitude with the required accuracy .
  The contribution of the Coulomb-hadron interference term in $dN/dt$
   depends on the parameters of the hadronic amplitudes and
  the Coulomb-hadron phase also depends on these parameters.
  After the normalization of the $dN/dt$ data we obtain 
    the differential elastic  cross  section.

   The analysis of the  elastic scattering data
  makes  several assumptions.
  Very often one assumes that
 the real and imaginary parts of the nuclear amplitude have the same
     $t$ exponential dependence;
  the cross-odd part of the scattering amplitude either has the same
   behavior of momentum transfer as
    the cross even part, or it is neglected;
 the  contributions of the spin-flip amplitudes are neglected.
  Note that the value of $\rho$ is heavily correlated with the normalization
  of $d\sigma/dt$. Its magnitude weakly impacts on the determination of
  $\sigma_{tot}$ only in the case where the normalization
  is known  exactly.

   There is no experiment which measures  the magnitudes of these
    parameters separately.
   Sometimes, to reduce experimental errors,
    the magnitude
    of some quantities is taken from another experiment.
   As a result, we can obtain a
    contradiction between the basic parameters for one energy  (for
    example, if we  calculate the imaginary part of the scattering
     amplitude, we can obtain a nonexponential behavior). It can lead to
      some errors in the analysis based on the dispersion relations.

    For example, let us take one experiment
     (there are  others which are very similar)
    which was carried out at $\sqrt{s}\ = 16. \  \div  \ 27.4 $ {\rm GeV}
   by the JINR-FNAL Collaboration \cite{kuznez}.
   To analyze   $dN/dt$ in this experiment,  the energy
  dependence of $\sigma_{tot}$ and $B$ were taken
  in  an analytical form supported
 by the  data of other experiments.
  But the new fit of the $dN/dt$ data with all free parameters gives
   values of $\rho$ which are systematically above the original 
  values \cite{selyf}.
   Such a conclusion was obtained early in \cite{Fajardo}.

    It means that the theoretical line obtained by  the dispersion relation
 will be higher than in the existing
  descriptions and close to a simple description with the use of only
 the crossing symmetric properties
   of the scattering amplitude (see, Figs.1,2).
  But it does not change essentially our prediction for
  high energies, only the value of the subtraction constant will be changed.

\section{Conclusions}

   The magnitudes of $\sigma_{tot}$, $\rho$ and slope - $B$ have to be
      determined in one experiment and determinations of their  magnitudes
      depend on each other.
  The procedure of
   extrapolation of the imaginary part of the scattering amplitude is
 significant for determining $\sigma_{tot}$.
   There can also exist some additional hypotheses
 which lead to a deviation of differential cross section 
at very small momentum transfer.
 For example, the  analysis of experimental data
 shows a possible manifestation of the hadron spin-flip amplitudes
at high energies.
The research into  spin effects
will be a crucial stone for different models and will help us
 understand the interaction and structure of particles, especially at large
distances.
All this raises the question about the measurement of spin effects
in  elastic hadron scattering at small angles  at
future accelerators.
Especially, we would like to note
the programs at RHIC
where the polarizations of both collider beams will be constructed.
  Additional information on  polarization data will help us
      only if we know sufficiently
     well the beam polarizations.
   So, the normalisations of $dN/dt$ and $A_N$ are most important
     for the determination of the magnitudes of these values.
     New methods of extracting the  magnitudes of these quantities
       are required.

\vspace{1.cm}

  {\it Acknowledgments.} I  would like to express my sincere thanks to
  the organizers   for the kind invitation and the financial support;
  and to  J.-R. Cudell and V. Ezhela for fruitful discussions.

\end{document}